\documentclass[twocolumn,aps,showpacs,prl,superscriptaddress]{revtex4} 

\usepackage{sidecap}
\usepackage{ulem}
\usepackage{epsfig}
\usepackage{amsmath,amssymb,amsthm}
\usepackage{graphicx}
\usepackage{bm}
\usepackage{color}

\DeclareMathAlphabet{\mathpzc}{OT1}{pzc}{m}{it}

\begin{document}

\renewcommand{\textfraction}{0.00}


\newcommand{\vAi}{{\cal A}_{i_1\cdots i_n}} 
\newcommand{\vAim}{{\cal A}_{i_1\cdots i_{n-1}}} 
\newcommand{\vAbi}{\bar{\cal A}^{i_1\cdots i_n}}
\newcommand{\vAbim}{\bar{\cal A}^{i_1\cdots i_{n-1}}}
\newcommand{\htS}{\hat{S}} 
\newcommand{\htR}{\hat{R}}
\newcommand{\htB}{\hat{B}} 
\newcommand{\htD}{\hat{D}}
\newcommand{\htV}{\hat{V}} 
\newcommand{\cT}{{\cal T}} 
\newcommand{\cM}{{\cal M}} 
\newcommand{\cMs}{{\cal M}^*}
\newcommand{\vk}{\vec{\mathbf{k}}}
\newcommand{\bk}{\bm{k}}
\newcommand{\kt}{\bm{k}_\perp}
\newcommand{\kp}{k_\perp}
\newcommand{\km}{k_\mathrm{max}}
\newcommand{\vl}{\vec{\mathbf{l}}}
\newcommand{\bl}{\bm{l}}
\newcommand{\bK}{\bm{K}} 
\newcommand{\bb}{\bm{b}} 
\newcommand{\qm}{q_\mathrm{max}}
\newcommand{\vp}{\vec{\mathbf{p}}}
\newcommand{\bp}{\bm{p}} 
\newcommand{\vq}{\vec{\mathbf{q}}}
\newcommand{\bq}{\bm{q}} 
\newcommand{\qt}{\bm{q}_\perp}
\newcommand{\qp}{q_\perp}
\newcommand{\bQ}{\bm{Q}}
\newcommand{\vx}{\vec{\mathbf{x}}}
\newcommand{\bx}{\bm{x}}
\newcommand{\tr}{{{\rm Tr\,}}} 
\newcommand{\bc}{\textcolor{blue}}

\newcommand{\beq}{\begin{equation}}
\newcommand{\eeq}[1]{\label{#1} \end{equation}} 
\newcommand{\ee}{\end{equation}}
\newcommand{\bea}{\begin{eqnarray}} 
\newcommand{\eea}{\end{eqnarray}}
\newcommand{\beqar}{\begin{eqnarray}} 
\newcommand{\eeqar}[1]{\label{#1}\end{eqnarray}}
 
\newcommand{\half}{{\textstyle\frac{1}{2}}} 
\newcommand{\ben}{\begin{enumerate}} 
\newcommand{\een}{\end{enumerate}}
\newcommand{\bit}{\begin{itemize}} 
\newcommand{\eit}{\end{itemize}}
\newcommand{\ec}{\end{center}}
\newcommand{\bra}[1]{\langle {#1}|}
\newcommand{\ket}[1]{|{#1}\rangle}
\newcommand{\norm}[2]{\langle{#1}|{#2}\rangle}
\newcommand{\brac}[3]{\langle{#1}|{#2}|{#3}\rangle} 
\newcommand{\hilb}{{\cal H}} 
\newcommand{\pleft}{\stackrel{\leftarrow}{\partial}}
\newcommand{\pright}{\stackrel{\rightarrow}{\partial}}

\title{LHC jet suppression of light and heavy flavor observables}

\date{\today}
 
\author{Magdalena Djordjevic}
\affiliation{Institute of Physics Belgrade, University of Belgrade, Serbia}
\author{Marko Djordjevic}
\affiliation{Faculty of Biology, University of Belgrade, Serbia}

\begin{abstract} 
Jet suppression of light and heavy flavor observables is considered to be an excellent tool to study the properties 
of QCD matter created in ultra-relativistic heavy ion collisions. We 
calculate the suppression patterns of light hadrons, D mesons, non-photonic 
single electrons and non-prompt $J/\psi$ in Pb+Pb collisions at LHC. We use a theoretical formalism that takes into account finite size {\it dynamical} QCD medium with finite magnetic mass effects and running coupling, which is integrated into a numerical procedure that uses no free parameters in model testing. We obtain a good agreement with the experimental 
results across different experiments/particle species. Our results show that the developed theoretical formalism can robustly explain suppression data in ultra relativistic heavy ion collisions, which strongly suggests that pQCD in Quark-Gluon Plasma is able to provide a reasonable description of the underlying jet physics at LHC.
\end{abstract}


\maketitle 

\section{Introduction} 

A major goal of RHIC and LHC experiments is to understand properties of a QCD medium created in ultra-relativistic heavy ion collisions. A powerful tool to map properties of such medium is to compare jet suppression~\cite{Bjorken} measurements with the corresponding theoretical predictions~\cite{Gyulassy,DBLecture,Wiedemann2013}.  Such predictions require accurate computations of jet energy loss, since the suppression is the consequence of the energy loss of high energy partons that move through the plasma~\cite{suppression,BDMS,BSZ,KW:2004}. In~\cite{MD_PRC,DH_PRL}, we developed a theoretical 
formalism for the calculation of the first order in opacity radiative energy 
loss in a realistic finite size dynamical QCD medium, which we subsequently generalized to the case of finite magnetic mass in~\cite{MD_MagnMass}. These studies, together with the calculations of the collisional energy loss is a finite size QCD medium that we previously developed~\cite{MD_Coll}, provide reliable framework for energy loss computation in Quark-Gluon Plasma (QGP). 

We here extend this formalism to running coupling and integrate it in a numerical procedure that can generate state of the art predictions for LHC experimental measurements. The  numerical procedure includes multi-gluon  fluctuations~\cite{GLV_suppress}, path length fluctuations~\cite{WHDG}  and most up-to-date jet production~\cite{Cacciari:2012,Vitev0912} and fragmentation functions~\cite{DSS}. Our strategy is to generate predictions for a diverse set of experimental probes, for which experimental data are available at LHC, in order to comprehensively test our understanding of QCD matter created in these collisions.  Specifically, we will generate suppression
predictions for light hadrons, D mesons, non-photonic single electrons and non-prompt $J/\psi$ in most central 2.76 TeV Pb+Pb
collisions at LHC. These predictions will be generated under the same numerical framework, by using the same set of parameters and with no free parameters used in model testing. Comparison of such predictions with the experimental measurements  allows testing to what extent pQCD calculations in QGP can explain the underlying jet physics at LHC. 
\section{Theoretical framework}

We use the generic pQCD convolution in order to calculate the quenched spectra of partons, hadrons, electrons and $J/\psi$:
\begin{eqnarray}
\frac{E_f d^3\sigma}{dp_f^3} &=& \frac{E_i d^3\sigma(Q)}{dp^3_i}
 \otimes
{P(E_i \rightarrow E_f )}\nonumber \\
&\otimes& D(Q \to H_Q) \otimes f(H_Q \to e, J/\psi). \; 
\label{schem} \end{eqnarray}
In the equation above $Q$ denotes quarks and gluons, while the terms in the equation correspond to the following: \smallskip
\\ {\it i)} $E_i d^3\sigma(Q)/dp_i^3$ denotes the initial quark spectrum, which is computed at next to leading order, according to~\cite{Cacciari:2012,FONLL} for charm and bottom quarks and according to~\cite{Vitev0912} for gluons and light quarks. \smallskip
\\ {\it ii)}  $P(E_i \rightarrow E_f )$ is the energy loss probability, which is generalized to include both radiative and collisional energy loss in a 
realistic finite size dynamical QCD medium, as well as multi-gluon~\cite{GLV_suppress} and path-length fluctuations~\cite{WHDG}; the path-length distributions for $0-5\%$ most central collisions are used as described in~\cite{Dainese}. Note that path-length distribution is the same for all jet varieties, since it corresponds to a geometric quantity. 
 \smallskip
\\ {\it iii)}
$D(Q \to H_Q)$ is the fragmentation function of quark or gluon 
$Q$ to hadron $H_Q$, where for light hadrons, D mesons and B mesons we use, respectively, DSS~\cite{DSS}, BCFY~\cite{BCFY} and KLP~\cite{KLP} fragmentation functions. \smallskip
\\ {\it iv)} For heavy quarks, we also have the decay of hadron $H_Q$ into single electrons or $J/\psi$, which is represented by the functions $f(H_Q \to e, J/\psi)$. The decays of D, B mesons to non-photonic single electrons, and decays of B mesons to non-prompt $J/\psi$ are obtained according to~\cite{Cacciari:2012}.  
\medskip

Furthermore, in the calculations of jet suppression, we use the following assumptions: \\ 
\smallskip
{\it i)} The final quenched energy $E_f$ is sufficiently large, so that we can employ the Eikonal approximation. \\ 
\smallskip {\it ii)}  The jet to hadron fragmentation functions are the same for $e^+e^-$ and Pb+Pb collisions, which is expected to be valid in a deconfined QCD medium. \\ 
\smallskip {\it iii)} We can separately treat radiative and collisional energy loss, so that we first calculate how quark and gluon spectrums change due to radiative energy loss, and then due to collisional energy loss. This approximation is reasonable as long as the radiative and collisional energy losses are sufficiently small (as assumed by soft-gluon, soft-rescattering approximation, which is employed in all energy loss calculations so far), and when collisional and 
radiative energy loss processes are decoupled from each other (as follows from HTL approach~\cite{RadVSColl} that is used in our energy loss 
calculations~\cite{MD_PRC,DH_PRL,MD_Coll,MD_MagnMass}). 
 \bigskip

Numerical method for including multi gluon fluctuations in the radiative energy loss probability is presented in Refs.~\cite{Djordjevic:2004nq,Djordjevic:2005db}. We recently generalized the procedure~\cite{MD_PRC2012}  to include the radiative energy loss in 
finite size dynamical QCD medium~\cite{MD_PRC,DH_PRL}, as well as finite magnetic mass effects~\cite{MD_MagnMass}. Specifically, the gluon radiation spectrum is extracted from Eq. (10) in~\cite{MD_MagnMass}. We furthermore approximate the full fluctuation spectrum in collisional energy loss probability  by a Gaussian with a mean determined by the average 
energy loss and the variance determined by 
$\sigma_{coll}^2 = 2 T \langle \Delta E^{coll}(p_\perp,L)\rangle 
$~\cite{Moore:2004tg,WHDG}; here, $\Delta E^{coll}(p_\perp,L)$ is extracted 
from Eq.~(14) in~\cite{MD_Coll}, while $T$, $p_\perp$ and $L$ are, respectively, the temperature of the medium, the initial jet momentum and the length of the medium traveled by the jet. 

We further extend this formalism by introducing the running coupling in the following way: In the radiative energy loss case, the coupling appears through the term $\mu_E^2 \, \alpha^2_S$~\cite{MD_MagnMass}. This can be factorized as
 $\mu_E^2 \, \alpha_S (Q_v^2) \, \alpha_S (Q_k^2)$, where the first $\alpha_S$ corresponds to the interaction between the jet and the virtual (exchanged) gluon, while the second $\alpha_S$ corresponds to the interaction between the jet and the radiated gluon (see~\cite{MD_PRC}). For the running coupling, Debye mass $\mu_E$ is obtained by self-consistently solving the following  equation~\cite{Peshier2006}:
\beqar
\frac{\mu_E^2}{\Lambda_{QCD}^2} \ln \left(\frac{\mu_E^2}{\Lambda_{QCD}^2}\right)=\frac{1+n_f/6}{11-2/3 \, n_f} \left(\frac{4 \pi T}{\Lambda_{QCD}} \right)^2,
\eeqar{mu}
where $\Lambda_{QCD}$ is perturbative QCD scale, and $n_f$ is number of the effective degrees of freedom. Running coupling $\alpha_S (Q^2)$ is defined as~\cite{Field} 
\beqar
\alpha_S (Q^2)=\frac{4 \pi}{(11-2/3 n_f) \ln (Q^2/\Lambda_{QCD}^2)}.
\eeqar{alpha}
To obtain $\alpha_S (Q_v^2)$, note that  $Q_v^2=E \, T$~\cite{Peigne2008}, where $E$ is the energy of the jet. Similarly, to obtain $\alpha_S (Q_k^2)$, note that the off-shellness of the jet prior to the gluon radiation $Q_k^2=\frac{\bk^2+M^2 x^2 + m_g^2}{x}$~\cite{MD_PRC}, where $\bk$ is transverse momentum of the radiated gluon, $M$ is the jet mass,  $x$ is the longitudinal 
momentum fraction of the jet carried away by the emitted gluon, and 
$m_g=\mu_E/\sqrt 2$ is the effective mass for gluons with hard momenta 
$k\gtrsim T$~\cite{DG_TM}. Note that, as introduced above, $\alpha_S (Q_{k, \,v}^2)$ are infrared safe (and moreover of a moderate value), so there is no need to introduce a cut-off in $\alpha_S (Q^2)$, as is usually done with running coupling elsewhere (see e.g. ~\cite{Zakharov2008,Buzzatti2013}).

In the collisional  energy loss case, the coupling appears through the term 
$\alpha^2_S$~\cite{MD_Coll}, which can be factorized as $\alpha_S (\mu_E^2) \, \alpha_S (Q_v^2)$~\cite{Peigne2008}, with  $\alpha_S (Q^2)$ given by Eq.~(\ref{alpha}).

\section{Numerical results} 

We here show our suppression predictions for light and heavy flavor observables in central 2.76 TeV Pb+Pb collisions at LHC. The following parameters are used in the numerical calculations: QGP of temperature $T{\,=\,}304$\,MeV (as determined by ALICE, see Ref.~\cite{Wilde2012}) with 
 effective light quark flavors $n_f{\,=\,}2.5$, and perturbative QCD scale of $\Lambda_{QCD}=0.2$~GeV. The light quark mass is assumed to dominated by the thermal mass $M{\,=\,}\mu_E/\sqrt{6}$, where Debye mass
$\mu_E \approx 0.9$~GeV is obtained by self-consistently solving  Eq.~(\ref{mu}). The values for the magnetic mass $\mu_M$ are taken in the range $0.4 \, \mu_E < \mu_M < 0.6 \, \mu_E$~\cite{Maezawa,Bak}; the gluon mass is  $m_g=\mu_E/\sqrt{2}$~\cite{DG_TM}, while the 
charm  and the bottom mass are, respectively, $M{\,=\,}1.2$\,GeV  and $M{\,=\,}4.75$\,GeV. Path-length distribution, parton production, fragmentation functions and decays, which are used in the numerical calculations, are specified in the previous section.

\medskip
\begin{figure*}
\epsfig{file=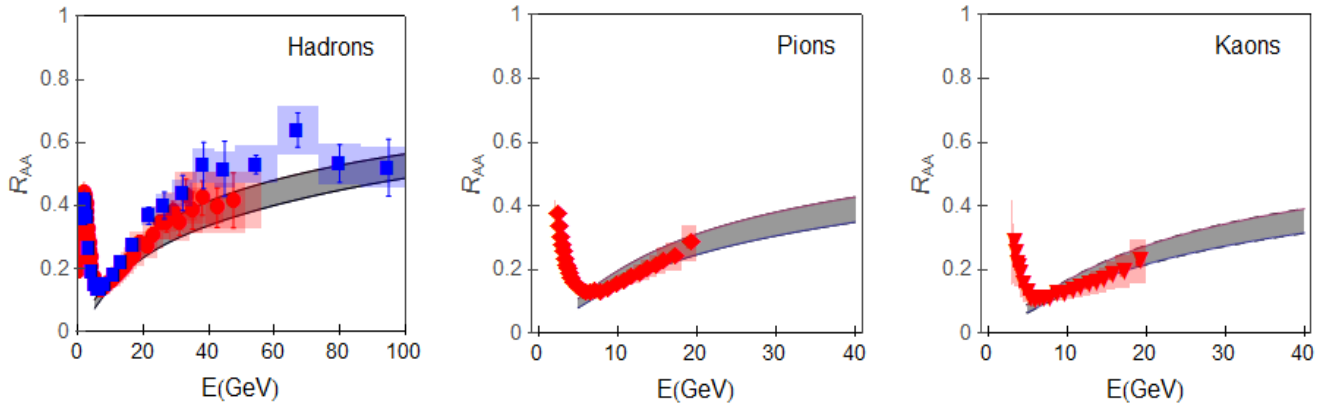,width=6.9in,height=2.3in,clip=5,angle=0}
\vspace*{-0.25cm}
\caption{{\bf Theory vs. experimental data for momentum dependence of light flavor $R_{AA}$.} The left panel shows the comparison of light hadron suppression 
predictions with experimentally measured $R_{AA}$ for charged particles. The red circles and the blue squares, respectively, correspond to ALICE~\cite{ALICE_h} and CMS~\cite{CMS_h} experimental data. The central panel shows the 
comparison of pion suppression predictions with preliminary $\pi^\pm$ ALICE~\cite{ALICE_preliminary1} $R_{AA}$ data (the red rhomboids), while the right panel shows the 
comparison of kaon suppression predictions with preliminary $K^\pm$ $R_{AA}$ ALICE data~\cite{ALICE_preliminary1} (the red triangles). All the data correspond to 0-5\% central 2.76 TeV Pb+Pb 
collisions. On each 
panel, the gray region corresponds to the case where $0.4 < \mu_M/\mu_E < 0.6$, with the upper (lower) 
boundary of each band that corresponds to $\mu_M/\mu_E =0.4$ ($\mu_M/\mu_E =0.6$).}
\label{LightHadronRaa}
\end{figure*}

\begin{figure*}
\epsfig{file=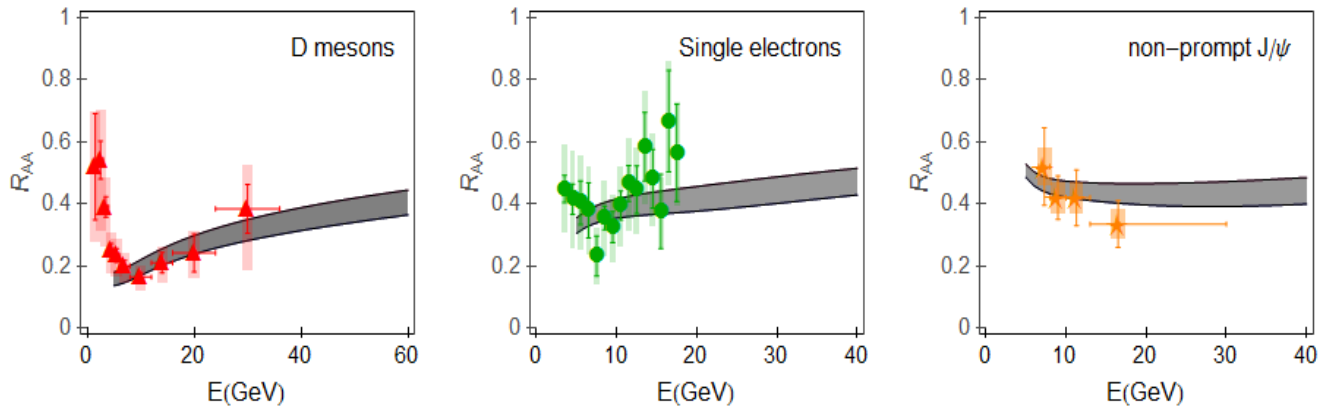,width=6.9in,height=2.3in,clip=5,angle=0}
\vspace*{-0.25cm}
\caption{{\bf Theory vs. experimental data for momentum dependence of heavy flavor $R_{AA}$.} The left panel shows the comparison of D meson suppression 
predictions with D meson $R_{AA}$  ALICE preliminary data~\cite{ALICE_preliminary2} (the red triangles) in 0-5\% central 2.76 TeV 
Pb+Pb collisions. The central panel shows the 
comparison of non-photonic single electron suppression  with the corresponding ALICE preliminary data~\cite{ALICE_SE} (the green circles)  in 0-10\% central  2.76 TeV Pb+Pb collisions. The right panel shows the comparison of $J/\psi$ suppression predictions with the preliminary non-prompt $J/\psi$ $R_{AA}$ CMS data~\cite{Mihee2013} (the orange stars) in 0-100\% 2.76 TeV 
Pb+Pb collisions. The gray region on each panel is as defined in Fig.~\ref{LightHadronRaa}. }
\label{HeavyHadronRaa}
\end{figure*}

The panel in Figure~\ref{LightHadronRaa} shows momentum dependence of $R_{AA}$ for light flavor observables, i.e. charged hadrons, 
pions and kaons at LHC. The predictions are 
compared with the relevant ALICE~\cite{ALICE_h,ALICE_preliminary1} and 
CMS~\cite{CMS_h} experimental data in central 2.76 TeV Pb+Pb collisions at 
LHC. The panel in Figure~\ref{HeavyHadronRaa}  shows momentum dependence of $R_{AA}$ for heavy flavor observables, i.e.  D mesons, 
non-photonic single electrons and non-prompt $J/\psi$ at LHC; here, the predictions 
are also compared with the corresponding  ALICE~\cite{ALICE_SE,ALICE_preliminary2} and 
CMS~\cite{Mihee2013} experimental data at 2.76 TeV Pb+Pb collisions at 
LHC. 

For all six observables shown in Figs.~\ref{LightHadronRaa} and~\ref{HeavyHadronRaa}, we see a very good agreement between the predictions and the 
experimental data. The left panel in Fig.~\ref{LightHadronRaa} (charged hadrons) shows an excellent agreement with ALICE data~\cite{ALICE_h} (the red circles), and a somewhat worse agreement with CMS data~\cite{CMS_h} (the blue squares), since CMS charged hadron $R_{AA}$ is systematically somewhat above the corresponding ALICE $R_{AA}$. Both the central and the right panel show excellent agreement between the theoretical predictions and preliminary ALICE  pion and kaon $R_{AA}$ data~\cite{ALICE_preliminary1}; note that these predictions reproduce a fine qualitative resolution between pion and kaon $R_{AA}$ data, i.e. the fact that observed kaon suppression is systematically somewhat larger compared to the pion suppression. For the heavy flavor measurements, predictions for D meson data (the left panel in Fig.~\ref{HeavyHadronRaa}~) show a similarly good agreement with the available experimental ALICE preliminary data~\cite{ALICE_preliminary2}. Though the preliminary non-photonic single electron data~\cite{ALICE_SE} are quite noisy (the central panel in Fig.~\ref{HeavyHadronRaa}), there is a very good agreement with the corresponding theoretical predictions;  further reduction of the error bars is needed for a clearer comparison. Finally, we also see a good agreement between the theoretical predictions and CMS preliminary non-prompt $J/\psi$ data~\cite{Mihee2013} (the right panel in Fig.~\ref{HeavyHadronRaa}), except for the last data point, for which the error bars are very large. Regarding $J/\psi$ data, one should here note that our predictions (which are done for the central collisions) are compared with the available 0-100\% centrality measurements; the change in the centrality is expected to increase the suppression compared to the results presented here, though based on~\cite{Mihee2013}, we expect that the increase will not be significant.

\section{Conclusions} 

A major theoretical goal in relativistic heavy ion physics is to develop a theoretical framework that is able to consistently explain both light  and heavy flavor experimental data. We here presented suppression calculations for light and heavy flavor observables, by taking into account finite size dynamical QCD medium with magnetic mass effects and running coupling taken into account. We generated predictions for six independent observables, i.e. light hadrons, D mesons, non-photonic single electrons and non-prompt $J/\psi$, for which experimental measurements at LHC are available. To our knowledge, a suppression predictions for such a diverse set of probes, generated by the same theory, within the same numerical procedure/parameter set, were not provided before. Furthermore, no free parameters were used in the model testing, i.e. the parameter values were fixed in advance according to the standard literature values. Comparing these predictions with the available experimental data shows a robust agreement across the whole set of probes. Such agreement, together with our previous study addressing the heavy flavor puzzle at RHIC~\cite{MD_PRC2012}, indicates that the developed theoretical formalism can realistically model the QCD matter created in ultra-relativistic heavy ion collisions, and that pQCD in Quark-Gluon Plasma is able to describe the underlying jet physics at LHC. 

\bigskip 
{\em Acknowledgments:} 
This work is supported by Marie Curie International Reintegration Grants 
within the $7^{th}$ European Community Framework Programme 
(PIRG08-GA-2010-276913 and PIRG08-GA-2010-276996) and by the Ministry of Science and Technological 
Development of the Republic of Serbia, under projects No. ON171004 and 
ON173052 and by L'OREAL-UNESCO National Fellowship in Serbia. We thank I. Vitev and Z. Kang for providing the initial light 
flavor distributions and useful discussions. We also thank M. Cacciari 
for useful discussion on heavy flavor production and decay processes. 
We thank ALICE Collaboration for providing the shown preliminary data, and M. 
Stratmann and Z. Kang for their help with DSS fragmentation functions.

\end{document}